# Parametric study of lumbar belts in the case of low back pain: effect of patients' specific characteristics

**Rébecca Bonnaire[1,4], Woo-Suck Han[2], Paul Calmels[3], Reynald Convert[4], Jérôme Molimard[2]**

[1] IMT Mines Albi, Campus Jarlard, 81013 Albi, France

[2] Mines Saint-Etienne, Université Jean Monnet, INSERM, UMR1059, SAINBIOSE, CIS-EMSE, 158 Cours Fauriel, 42023 Saint-Etienne, France

[3] Laboratory of Exercice Physiology (LPE EA4338), University Hospital of Saint-Etienne, Hôpital Bellevue, 42055 Saint-Etienne, France

[4] Thuasne, 27 rue de la Jomayère, 42032 Saint-Etienne, France

***Abstract*** *Objective:* A numerical 3D model of the human trunk was developed to study the biomechanical effects of lumbar belts used to treat low back pain.

*Methods:* This model was taken from trunk radiographies of a person and simplified so as to make a parametric study by varying morphological parameters of the patient, characteristic parameters of the lumbar belt and mechanical parameters of body and finally to determine the parameters influencing the effects of low back pain when of wearing the lumbar belt. The loading of lumbar belt is modelled by Laplace's law. These results were compared with clinical study.

*Results:* All the results of this parametric study showed that the choice of belt is very important depending on the patient's morphology. Surprisingly, the therapeutic treatment is not influenced by the mechanical characteristics of the body structures except the mechanical properties of intervertebral discs.

*Discussion:* The numerical model can serve as a basis for more in-depth studies concerning the analysis of efficiency of lumbar belts in low back pain. In order to study the impact of the belt's architecture, the pressure applied to the trunk modelled by Laplace's law could be improved. This model could also be used as the basis for a study of the impact of the belt over a period of wearing time. Indeed, the clinical study shows that movement has an important impact on the distribution of pressure applied by the belt.

# 1 Introduction

Low back pain is a frequent disease which induces high cost to society in developed countries. So, from prevention to disability, a lot of treatments are proposed. Low back pain is clinically described by physio-pathological characteristics and its duration, from acute to chronic form [1]. One of the treatments for low back pain is the use of a lumbar orthotic device. The objectives of this device are to reduce mobility, decrease pain [2] and medication used which could induce some risks of iatrogenic complications.

The clinical efficacy of lumbar belt has been shown in the event of subacute and chronic low back pain [2]. However, few studies exist that clearly elucidate all the mechanisms of action proposed for this orthotics. After these studies, we could conclude that lumbar orthoses:

- limit the motion range of the trunk [3][4][5];
- reduce disc stresses by increasing the abdominal pressure [6][7][8];
- correct the posture by modifying lumbar lordosis [9][10][11];
- have no negative impact on muscle strength [12][13][14][15];
- have muscle-relaxing effects [16][17][18][19];
- facilitate trunk mobility for daily living activities [2][18].

No overall study of the mechanisms of action of lumbar belts and no study based on numerical modelling of their effects on the trunk to reduce low back pain were found through the review of the literature.

However, numerous models exist for the human trunk and, more specifically, the spine. These models can be divided into three types: detailed models, simplified models, and hybrid models, i.e. containing both highly detailed zones and highly simplified zones. Numerical models have been used, for example, to compare the influence of parameters for surgical treatments or other techniques [20][21][22], to analyze the deformation of the spine ([23][24][25]) or the effect of carrying a load ([26][27]). One study on the mechanisms of action of rigid orthosis in the treatment of scoliosis appears to be similar to the present study and report that biomechanical analysis using computational modelling could be useful in the design of rigid orthosis [28].

In this Finite Element Analysis (FEA), a lumbar orthosis, named lumbar belt, was studied. This orthosis is a common treatment of subacute or chronic low back pain. The main objective of this study is to evaluate the parameters that can influence the choice of lumbar belts types to treat low back pain, using a finite elements model. The studied parameters will be related to the patient's morphology, the belt and the mechanical characteristics of the human trunk.

## 2 Methodology

The aim is to perform a screening of the factors influencing the global mechanical effect of the belt. For this purpose, a simplified numerical model was built so as to change easily trunk geometry according to different typical patients. Unfortunately, this induces a compromise between computing time and precision on the results. The model might not render properly local effects but must be reliable on global ones. Indeed, a special attention must be done on the spine modeling, because pain is supposed to be caused from intervertebral discs. The geometry is simplified on trunk and vertebrae. Finally, the mechanical behavior is assumed to be linear and a simple Laplace law simulates the pressure applied by the belt. All these assumptions make a parametrical study possible, even if they track important limitations.

### 2.1 Trunk modelling

#### 2.1.1 Construction of geometric model

A simplified 3D model of the human trunk was developed for FEA. This model is built using frontal and sagittal radiographies of the trunk of one person. Using these radiographies, the following elements are measured:

- the lateral inclination of each intervertebral endplate in the sagittal plane,
- the height at their center of the thoracic and lumbar vertebrae,
- the diameter of the vertebral anterior segment at thoracic and lumbar level,
- the height at their center of the intervertebral discs in the thoracic and lumbar regions,
- the width of the trunk at the chest, below the chest, at the waist and hips in the frontal plane,
- the thickness of the trunk at the chest, below the chest, at the waist and hips in the sagittal plane.

Each vertebra is modelled as a cylinder based on the radiographs. Lumbar lordosis and thoracic kyphosis are constructed using two circle arcs obtained thanks to the lateral inclinations of each vertebral endplate. It is assumed that the vertebra have parallel lower and upper endplates. The posterior segment of vertebra is not represented in this model since their action is assumed to be negligible for this study. The intervertebral discs are also modelled as a cylinder having two distinct parts: a central circular part corresponding to 30% of the volume representing the nucleus and a peripheral part representing the annulus [29].

The tissues composing the trunk, excluding the spinal column, are divided into three zones in order to represent the abdomen, the iliac crests and one other part including the thorax and the soft tissues in the back zone (see Fig. 1). This geometric model is simplified to easily change the trunk morphology in the simulation and realize the parametric studies. The surface of the trunk, corresponding to the skin, has different mechanical properties from the rest of the volume. This skin has a thickness of 4 mm; the value was determined by ultrasound on the trunk of two human subjects (see Fig. 2).

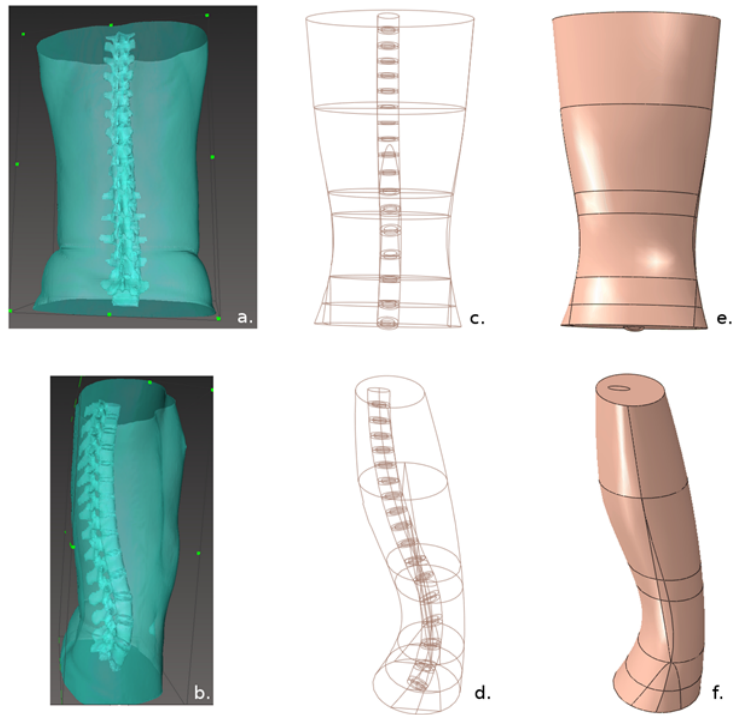

**Fig. 1.** Representation of modelling of trunk geometry for this study: trunk geometry before modelling a. from the back b. in profile; modelled trunk geometry c. and e. from the back, d. and f. in profile. Figures a and b are free of copyright from: https://www.osirix-viewer.com/re-sources/dicom-image-library/.

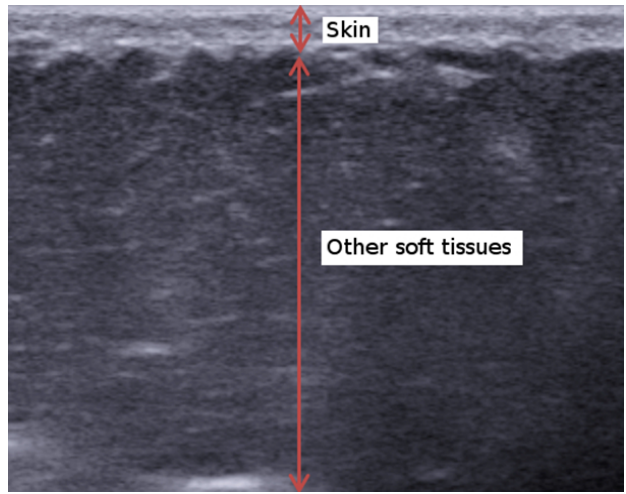

**Fig. 2.** Ultrasound image used to measure the thickness of the skin at the trunk. This is taken on one of co-authors by himself.

### 2.1.2 Finite Element modelling of the trunk

The geometry is meshed using ten-node tetrahedral second-order elements (C3D10H) using ABAQUS. Figure 3 shows a meshed model of the trunk. To determine the appropriate number of elements, a convergence study was done. This study consists in determining the mean abdominal and intradiscal pressure in the lumbar region as a function of the mean size of the elements. This study was carried out on the eight subjects chosen in this study, described in the next paragraph 2.2.

For the purpose of ensuring model simplicity, all the structures have elastic linear mechanical properties. As this study is a first approach, it would be interesting to modify this hypothesis in a future work. Mechanical properties can vary as a function of the age and musculature of the subject modelled. They are therefore considered as design parameters during this study.

After studies on different sets of boundary conditions, these were determined in order to obtain realistic movement of the trunk under the loading of the belt. When a patient wears a belt, the pelvis can rotate slightly relative to the transverse axis. In addition, the patient tends to straighten up. It was therefore decided to block the upper plate of the trunk in translation in the transverse plane. The top of the trunk can translate in the frontal plane. The lower plate of the trunk is restricted in order to enable rotation of this plate only in the transverse axis. This restriction is obtained by embedding the two points at the intersection of the ellipse, representing the extremity of the lower plate of the trunk, and the frontal plane of the trunk. The boundary conditions are represented in Fig. 3.

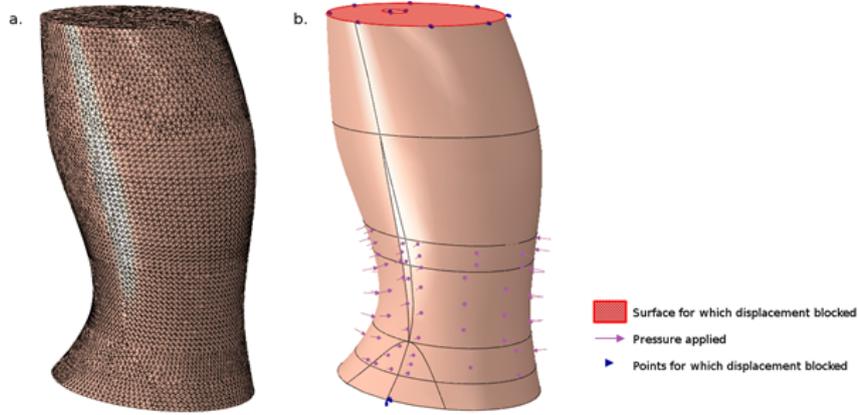

**Fig. 3.** Representation: a. of meshing of the finite trunk element model and b. of conditions at the limits.

### 2.1.3 Modelling of the lumbar belt

Lumbar belts are medical devices composed primarily of fabric in whom four steel whales are inserted in the dorsal part and a flexible whale in the front part. They may consist of one or several layers of fabric. In this finite element model, it was opted to represent the lumbar belt globally as application of pressure to the lumbar region of the trunk. The bottom of the belt is located at the lower endplate of lumbar vertebra L4. This pressure is obtained by the following Laplace's law [30]:

$$P = \frac{T}{R} \, if \, T \wedge R > 0 \tag{1}$$

$$P = 0 \, if \, T \wedge R \leq 0$$

where T is the linear tension of the belt on the trunk and R the radius of curvature of the trunk at the pressure application point. The linear pressure of the lumbar belt used in this study was obtained by tensile tests on the belts [31]. This overall belt model does not enable to simulate specific actions of certain structures of the belt, such as the back stays, but provides an adequate simplified model for the purposes of this study.

### 2.1.4 Assessment of belt modelling

In parallel of this simulation, a clinical study was carried out on fifteen low back pain subjects [31]. The interface pressure between the belt and the trunk was measured using piezoresistive pressure mapping system of 32×12 sensors [32]. The mean pressure measured during the clinical study could therefore be compared to the pressure by Laplace's law.

## 2.2 Parametric study

### 2.2.1. Input parameters

To show how lumbar belt wearing influences mechanical conditions, a numerical design of experiments was built. Its input parameters are:

- the morphological parameters of the patient: height, corpulence and lumbar lordosis angle,
- the characteristic parameters of the lumbar belt: belt type and height,
- the mechanical properties of the structures composing the model: skin, abdomen, annulus, nucleus, bone and others soft tissues.

To vary the patient's morphological parameters, a basic geometry was constructed as described in the previous paragraph 2.1 "Trunk modelling". This geometry was obtained from radiographies of one patient, without 3D deformation of the spine. This geometry was proportionally modified in order to obtain two trunk's geometries: one corresponding to a person with the height of 1.60m and one corresponding to a person with the height of 1.80m. The basic geometry corresponded to a person with a small waist circumference. This basic geometry was therefore considered as "slim". The trunk thicknesses in the sagittal plane at the waist and hips levels were doubled in order to obtain the geometry of a "fat" person. The inclination angles of the lumbar vertebrae were modified to obtain subjects with a low lumbar lordosis of 33° or a high lumbar lordosis of 53°. Subjects with hypo- and hyper-lordosis could therefore be modelled. Finally, eight characteristic geometries were obtained by varying the height (tall/small), stature (slim/large) and lumbar lordosis (hypo/hyper-lordosis).

Laplace's law modeled two types of lumbar belts. The first one has a single fabric thickness (LombaSkin®) and the other is more rigid having a posture-correcting strap (Lombacross Activity®). Each belt type has two different heights: 21 cm and 26 cm.

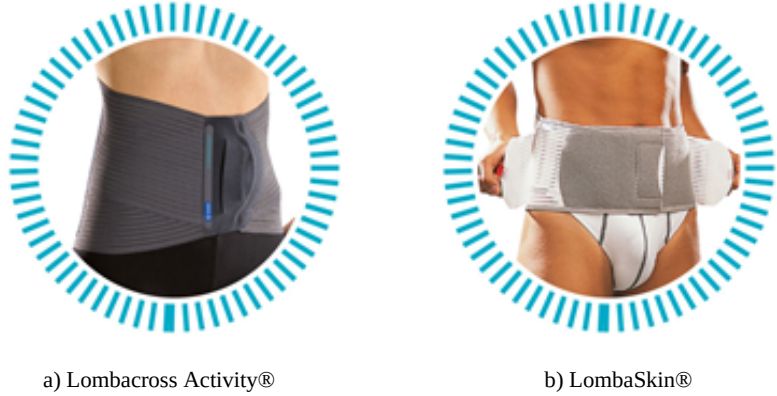

a) Lombacross Activity®          b) LombaSkin®

**Fig. 4.** Examples of lumbar belts used in this study provided by Thasne®: a) Lombacross Activity®, b) LombaSkin®.

With linear elastic assumptions, we only need Young's moduli and Poisson's ratios to describe mechanical behavior of materials. Poisson's ratios, used in the model and taken from the literature [28][33][34][35], are shown in Table 1. Only Young's moduli will vary in this parametric study. They are defined by continuous parameters, in contrast with the other parameters, and may vary within a range on the basis of the data in the literature [28][33][34][35].

This parametric study therefore includes 11 input parameters (see Table 2).

### 2.2.2 Output parameters

The following three output parameters were observed:

- the variation in abdominal pressure, since it characterizes the direct impact of belts on the trunk,
- the variation in lordosis, characterized by measurement of the mean displacement of the lumbar vertebrae in the sagittal plane to determine the impact on the spine posture,
- the variation in intradiscal pressure in the lumbar spine.

### 2.2.3. Design of experiments

The design method used is a stratified 100-experiment Latin hypercube design. It was taken since it makes possible to cover the entire study space with a relatively small number of experiments and at the same time, enabling binary or continuous input parameter analysis.

**Table 1.** Poisson's ratio for the different components in the model.

| Model structure | Poisson's ratio |
|---|---|
| Abdomen | 0.45 |
| Annulus | 0.45 |
| Nucleus | 0.48 |
| Skin | 0.2 |
| Bone | 0.3 |
| Other soft tissues | 0.45 |

**Table 2.** Morphology of the subjects modelled, characteristics of the belts studied and variation range for the mechanical properties of the structures for the parametric study

| Parameters | -1 | 1 |
|---|---|---|
| Subject's height (cm) | 160 | 180 |
| Subject's corpulence | slim | large |
| Subject's lordosis (°) | 33 | 53 |
| Belt type | LombaSkin® | LombaCross Activity® |
| Belt height (cm) | 21 | 26 |
| Young's modulus of the abdomen (MPa) | 0.01 | 1 |
| Young's modulus of the annulus (MPa) | 4 | 14.9 |
| Young's modulus of the nucleus (MPa) | 3 | 8 |
| Young's modulus of the skin (MPa) | 1 | 5 |
| Young's modulus of the bone (MPa) | 1,000 | 12,000 |
| Young's modulus of the other soft tissues (MPa) | 0.55 | 1 |

## 3 Results

### 3.1 Convergence study

Fig. 5 represents the variation in mean abdominal pressure as a function of the size of the elements for the most slim and smallest patient with hypolordosis in the study. Convergence is obtained for an element size of less than 15 mm, corresponding to a little over 80,000 elements in this case. Similar results were found for the other seven geometries modelled. The convergence study done with the eight used subject geometries finally shows that convergence is achieved for all the cases when the elements have a mean size of 10 mm. This mean size corre-sponds to models including 242,230 to 481,850 elements. It was therefore selected to perform the numerical design of experiments.

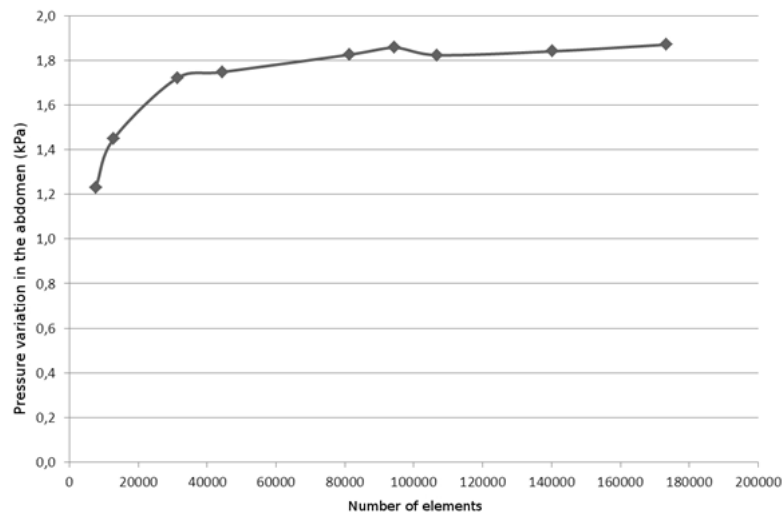

**Fig. 5.** Result of the convergence study: variation in mean abdominal pressure in the model as a function of the number of elements.

### 3.2 Assessment of belt modelling

In the clinical study, the mean pressure applied is 1.12 kPa or 2.03 kPa, respectively, for a single or double fabric thickness lumbar belt. In our model, the mean pressure applied to model by the belt is 1.06 kPa or 2.16 kPa, respectively, for a single or double fabric thickness lumbar belt. The mean error is about 6% and resulting from simplified modelling of the belt by Laplace's law. The belt model is acceptable.

### 3.3 Example of detailed results for a modelled subject

Fig. 6 shows the pressure and displacement results for two subjects modelled in this study.

The displacements are very different between the subject's abdomen and back. However, the displacements are symmetrical in the sagittal plane. At the iliac crests, displacements are almost zero.

The pressure inside elements is the highest at the iliac crests. This pressure is also high in the abdomen. It is negative in the lumbar region.

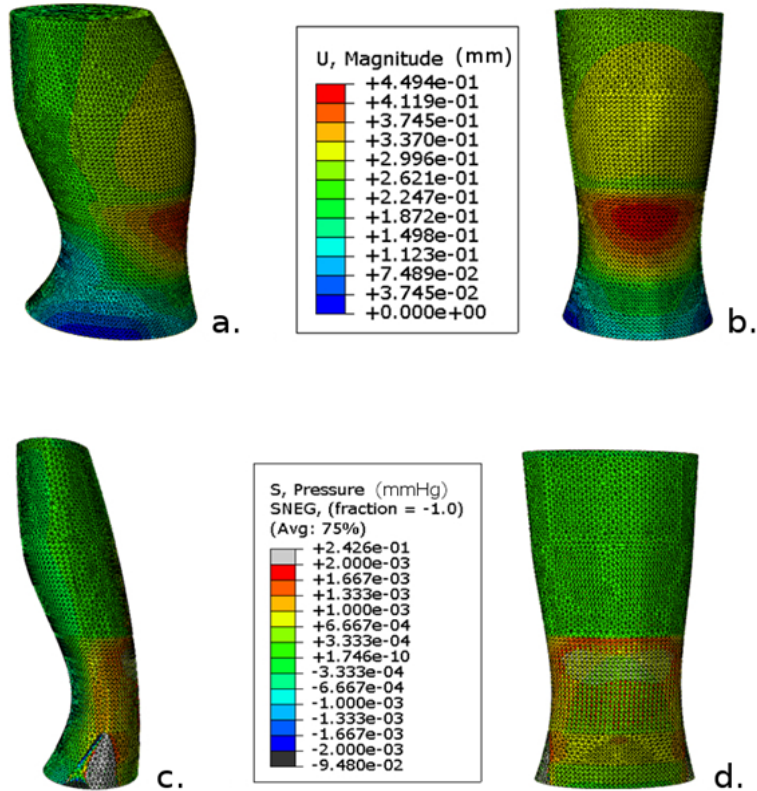

**Fig. 6.** Displacement result for a fat, tall subject with hyperlordosis applied by Lombacross Activity®: a. in profile and b. from the front, and pressure result inside element for a slim and small subject with hypolordosis: c. in profile and d. from the front.

### 3.4 Parametric study

The stratified Latin hypercube design made it possible to construct a linear model according to Equation 2:

$$Y = \beta_0 + \beta_s X_s + \beta_c X_c + \beta_l X_l + \beta_{bt} X_{bt} + \beta_h X_h + \beta_{pab} X_{pab} + \beta_{pan} X_{pan} + \beta_{pn} X_{pn} + \beta_{ps} X_{ps} + \beta_{pb} X_{pb} + \beta_{ot} X_{ot}$$

(2)

where:

- - Y = the output variable, which may be either the variation of abdominal pressure or the variation of intradiscal pressure, or the mean displacement of the lumbar vertebrae in the sagittal plane,
- - $\beta_0$ = the mean value of the linear model,

- - $\beta_{i=}$ the linear coefficients of the model and have an index $s$ for the patient's height, $c$ for the patient's corpulence, $l$ for the patient's lordosis, $bt$ for the belt type, $h$ for the belt height, $pab$ for Young's modulus of the abdomen, $pan$ for Young's modulus of the annulus, $pn$ for Young's modulus of the nucleus, $ps$ for Young's modulus of the skin, $pb$ for Young's modulus of the bone and $ot$ for Young's modulus of the other soft tissues,
- - $X =$ is the input variable with the same indices as for the $\beta$ coefficients.

X is reduced in a [-1 +1] range. The mean displacement of the vertebrae is taken to be positive when the lordosis increases. The model chosen was a linear one without interaction since the interaction parameters are weak compared to the main parameters.

Fig. 7 plots the linear coefficients of the model for the variation in abdominal and intradiscal pressures and the mean displacement of the lumbar vertebrae, related to the variation in lordosis. Because of the input parameters reduction to [-1 +1], this graphic representation of the linear model developed enables to determinate the parameters with the greatest influence. It shows that greater is the linear coefficient compared to the mean value of the output parameter, greater is the influence of the input parameter corresponding to this linear coefficient.

Hence the parameters with the greatest influence on displacement of the lumbar vertebrae are waist circumference, belt height and type, and lumbar lordosis before wearing the belt; the parameters with the least influence are the subject's height and Young's modulus of the annulus.

With respect to the variation in abdominal pressure, the parameters with the greatest influence are belt height and type as well as lumbar lordosis; the parameters with the least influence are the waist circumference and the mechanical properties of the skin and the annulus.

Finally, the parameters with the greatest influence to the variation in intradiscal pressure are the belt type, the subject's height and waist circumference, while the parameter with the least influence is Young's modulus of the skin.

A summary of the influence of input parameters on output parameters is presented in Table 3.

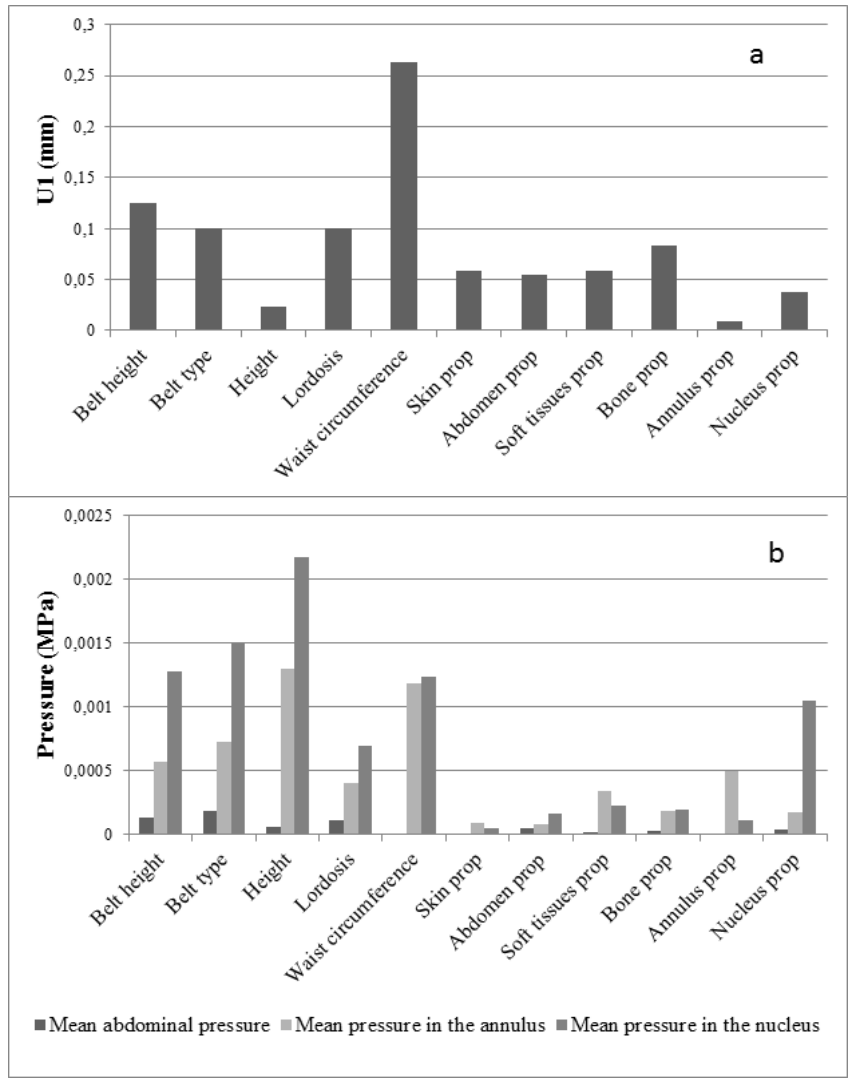

**Fig. 7.** Representation of the linear coefficients obtained in absolute values according to the analysis by experimental design for: a. displacement of the lumbar vertebrae, b. variation in abdominal and intradiscal pressures.

**Table 3.** Influence of input parameters on output parameters.

|  | Mean abdominal pressure | Mean intradiscal pressure | Variation in lumbar lordosis |
|---|---|---|---|
| Belt height | ++ | + | + |

| Belt type | ++ | ++ | + |
|---|---|---|---|
| Subject's height | + | ++ | + |
| Lumbar lordosis | ++ | + | + |
| Subject's waist circumference | -- | ++ | ++ |
| Mechanical properties of the skin | -- | -- | - |
| Mechanical properties of the abdomen | + | - | - |
| Mechanical properties of the soft tissues | - | + | - |
| Mechanical properties of the bone | - | - | - |
| Mechanical properties of the annulus | -- | + | -- |
| Mechanical properties of the nucleus | + | + | - |

## 4  Discussion

### 4.1  Trunk model

The developed FE trunk model is geometrically very simple but includes numerous structures, such as the abdomen, iliac crests and spine. Its main advantage lies in the fact that it enables rapid modelling of different morphologies solely from two complete radiographies of the trunk. This model can also be adapted to only simulate part of the trunk, such as the lumbar region. Another advantage of this model is a reliable calculation time, which means that it is possible to test numerous configurations in a reasonable computing time: a few hours for one configuration on a 2-core computer.

During the construction of this model, several hypotheses were made.

Firstly, deformations were assumed to be small, making it possible to model the structures using elastic linear properties. This hypothesis is confirmed since strains less than 1% for each component make it possible to stay in the linear zone of the biomechanical structures behavior.

Secondly, in order to simplify the model and not to model the posterior parts of the vertebrae present in other models [36], it was assumed that rotation of the ver-tebrae in the transverse axis is small on application of the pressure representing the belt wearing. The rotation of the vertebrae observed in the model is less than 5°. This rotation is small compared to the lumbar lordosis and thoracic kyphosis of the patients prior to application of pressure. This second hypothesis is also validated therefore.

Thirdly, the initial stresses in the trunk due to gravity and the presence of the head and legs are not taken into account in the model, in contrast with other models [28][37][38]. This hypothesis is correct since the study only concerns the pressure variation and displacement of vertebrae due to the belt wearing and not absolute pressure or deformation values.

Fourthly, it was opted to represent the action of the lumbar belt by a pressure range obtained by Laplace's law, as has already been proposed for compression socks in another study [30]. The mean pressures applied by lumbar belts obtained in a clinical study in fifteen subjects with low back pain [31] demonstrate that this simplistic model is acceptable. The difference in the interface pressure applied by the lumbar belt between simulation and clinical measurement is only 6%.

In the context of our study, the hypotheses adopted are therefore all justified. In addition, the clinical study demonstrates that the variation in trunk volume measured in humans during wear of a lumbar belt is almost identical to the variation in volume obtained in this model. This proves that, for an overall study of the mechanism of action of lumbar belts, our model is valid.

## 4.2 Mechanism of action of a lumbar belt

Observation of the pressure and displacement variations on a patient can explain some mechanisms of therapeutic action of lumbar belts. The belt applies pressure on the skin, and greatly increases abdominal pressure. This increase leads to a slight change in lordosis and to reduce constraints on vertebrae and intervertebral structures. The variation in intradiscal pressure is low, at around 1%. This low-pressure variation is close to pressure variation obtained in other studies on rigid orthotics [39].

Pressure is the highest in the iliac crests. It can be explained by the fact that the iliac crests are highly rigid structures compared to the soft tissues of the trunk. High pressure is linked to the feeling of well-being. If the pressure applied by the belt is too high, the patient will find it more difficult to accept his/her belt [40].

The parametric study made it possible to determine some parameters influencing the variation in abdominal and intradiscal pressure and displacement of the lumbar vertebrae, related to the modification in lumbar lordosis, which can play an important role in pain relief. These three output parameters are complementary to contribute to the therapeutic effects of lumbar belts in the treatment of low back pain.

Finally, the results of this parametric study raise the following points with respect to the clinical use of lumbar belts in the treatment of low back pain:

- the choice of belt is very important;
- it is necessary to take into account the patient's morphology;
- the belt treatment is independent of mechanical properties of soft tissue, excluding the mechanical properties of intervertebral discs.

As a matter of fact, all the effects of a lumbar belt cannot be described by the model. In particular, pressure induces proprioceptive effects on the body and an active change in the posture. This approach only shows a passive effect of the belt on the trunk.

## 5  Conclusion

Using a simplified model of the human trunk obtained on the basis of two radiographies, a parametric study made it possible to determine some parameters influencing the efficiency of lumbar belt in the therapeutic treatment of low back pain. These parameters are the patient's morphology and the type of lumbar belt.

The numerical model used is a simple one. It can serve as a basis for more indepth studies concerning the analysis of efficiency of lumbar belts in low back pain. In order to study the impact of the belt's architecture, modelling of the pressure applied to the trunk using Laplace's law could be improved. For example, the linear tension used in Laplace's law could vary depending on the application zone. This linear tension could be higher at the dorsal part of the trunk, to reflect the presence of the back whales. For a more precise solution, it would be possible to model the belt using a volume structure including the different existing parts, i.e. the fabrics, the back and front whales, etc. This numerical model could also be used as the basis for a study of the impact of the belt over a period of wearing time. Indeed, the clinical study shows that movement has an important impact on the distribution of pressure applied by the belt. This dynamic study would require modelling of the back muscles using volumetric structures with specific dynamic characteristics or beams.